\documentclass[12pt]{article}
\usepackage{amssymb,cite,epsf}
\newcommand{\tr}{{\rm Tr}}
\newcommand{\str}{\Sigma {\rm Tr}}

\renewcommand{\theequation}{\thesection.\arabic{equation}}
\begin{document}

\renewcommand{\thefigure}{\arabic{figure}}
\setcounter{figure}{0}
\begin{titlepage}
\begin{center}
{\hbox to\hsize{hep-th/9907088   \hfill  UCSD/PTH-99-10}}

\vspace{2cm}

\bigskip

{\Large \bf  Correlators of Short Multi-Trace Operators \\
             in ${\cal N}{\bf =4}$ Supersymmetric Yang-Mills }

\bigskip

\bigskip

\bigskip

{\bf Witold Skiba }\\

\smallskip

{\small \it Department of Physics, University of California at San Diego,\\
 La Jolla,  CA 92093}

\bigskip

{\tt  skiba@einstein.ucsd.edu}

\bigskip

\vspace*{1cm}
{\bf Abstract}\\
\end{center}
Multi-trace scalar operators in the $(0,k,0)$ representations of $SU(4)_R$
share many properties with their single-trace analogs. These multi-trace
operators are primary fields of short representations of the ${\cal N}=4$
superconformal group. Thus, they have protected dimensions.
We show that two- and three-point correlator functions of such
operators do not receive radiative corrections at order $g^2$ in
perturbation theory for an arbitrary gauge group.

\bigskip

\bigskip

\end{titlepage}

\section{Introduction}
\setcounter{equation}{0}
\setcounter{footnote}{0}

Short representations of ${\cal N}=4$ superconformal algebra play an
important role in the correspondence between supersymmetric Yang-Mills
(SYM) theory and supergravity in anti-de Sitter (AdS)
space~\cite{Maldacena,GKP,Witten}. Operators in the short representations
have protected dimensions, while the dimensions of the long representations
depend on the coupling constant. Short operators with a single trace
over the gauge indices are believed to correspond to single-particle states
of supergravity in the AdS background. Multi-trace operators relate
to multi-particle states. The long operators with single traces correspond
to string states.

Much work has been devoted to the investigations of correlator functions
using either the supergravity dual \cite{GKP,Witten,MIT,LMRS} or directly
computing in perturbation theory in SYM~\cite{ours,HW,G-RKP,G-RPS}.
For additional references see the review article~\cite{review}.
Since the supergravity calculations are applicable for large values of the
't~Hooft coupling, they cannot be easily compared with the weak-coupling
perturbation theory. Therefore, it is interesting to study non-renormalized
quantities. If a free-field value of some quantity coincides with its
large-coupling value obtained by supergravity calculations, then such a
result can be regarded as a hint for a non-renormalization theorem. For
example, two- and three-point functions of single-trace short operators
were found to  be equal to their free-field values~\cite{LMRS}.
By checking if such non-renormalization theorems hold at low orders of
perturbation theory one can test the conjectured duality. Perturbative
calculations in field theory also provide information beyond the leading
order in the $1/N$ expansion. See Refs.~\cite{ours,HW,G-RKP,Kovacs} for
calculations of two- and three-point functions. Of course, there should
be a symmetry reason for a non-renormalization theorem to hold to all
orders of perturbation theory and even
non-perturbatively~\cite{bonus1,bonus2,harmonic}.

In addition to single-trace short multiplets there are also
multi-trace operators in short representations of the ${\cal N}=4$
superconformal group. This fact is certainly known, but not widely
appreciated. See Refs.~\cite{Chalmers,Bianchi} for recent discussions
on multi-trace operators. The main result of this paper is that two-
and three-point functions of such short multi-trace 
operators are not renormalized at order $g^2$ in perturbation theory,
independent of the gauge group. Multi-trace operators
are in short representations whenever they transform in the same
representations of the flavor symmetry as the single-trace short operators.
That is, a primary operator proportional to $k$ powers of the scalar
field is short if it transforms in the $k$-index traceless symmetric 
representation of $SO(6)_R$, or equivalently in $(0,k,0)$ of $SU(4)_R$.
Two facts underline the relevance of short multi-trace operators for
the Maldacena duality. First, single- and multi-trace operators
have non-vanishing two-point functions already at the tree level.\footnote{
I thank Dan Freedman for pointing this out to me.}
Second, the supergravity dual is blind to the gauge structure of
${\cal N}=4$ SYM theory since only the gauge-invariant operators
enter the correspondence.
It is therefore possible that at sub-leading orders in $N$, the true
correspondence between the single-particle states of supergravity and
the short representations of superconformal ${\cal N}=4$ theory involves
a linear combination of single- and multi-trace operators. After all,
such operators have identical quantum numbers.

We first review why multi-trace operators in the $(0,k,0)$ representations 
of $SU(4)_R$ are indeed short and why their dimensions are protected.
Next, we show that two-point functions of arbitrary linear combinations
of operators in the $(0,k,0)$, either single- or multiple-trace are
not corrected at order $g^2$ in perturbation theory. Our results are an
extension of Ref.~\cite{ours}, we show that the techniques developed
there are also applicable to multi-trace operators. We follow the
notation of Ref.~\cite{ours}, where it was shown that two- and three-point
functions of short single-trace operators are not renormalized
at $O(g^2)$.
Afterwards, we repeat the calculation for a particularly simple case of
a three-point function, leaving for the Appendix the proof that arbitrary
three-point functions of linear combinations of short operators are not
renormalized at $O(g^2)$.

${\cal N}=4$ supersymmetry has 16 supercharges. Thus, a generic multiplet
is obtained by acting on a primary field with up to 16 supersymmetry
transformations. The on-shell transformations are~\cite{Brink}
\begin{eqnarray}
\label{susytransf}
  \delta A_{\alpha \dot \alpha}&=& \overline \eta^{I\dot \beta}
    \epsilon_{\dot \alpha \dot \beta} \psi_{I\alpha} + \eta^\beta_I
    \epsilon_{\alpha \beta} \overline \psi^I_{\dot \alpha}, \nonumber \\
  \delta \phi_{[IJ]}&=& \eta^\alpha_{[I} \psi_{J]\alpha} + \epsilon_{IJKL}
     \overline \eta^{K\dot \alpha} \overline \psi^L_{\dot \alpha}, \\ 
  \delta \psi_{I\alpha}&=& \eta^\beta _IF_{(\alpha \beta )}
     +\eta^\beta_J \epsilon_{\alpha \beta} [\phi_{IK},\phi^{JK}]+
       \overline  \eta^{J\dot \beta} D_{\alpha \dot \beta}\phi _{IJ},
       \nonumber \\
  \delta F_{(\alpha \beta )}&=& \overline \eta^{I\dot \gamma }
     D_{\dot \gamma (\alpha } \psi_{\beta )I}, \nonumber
\end{eqnarray}
where $\phi_{[IJ]}$, $\psi_{I\alpha}$, and $A_{\alpha \dot \alpha}$ are
the scalar, fermion, and gauge components of the ${\cal N}=4$ multiplet,
respectively. $F_{(\alpha \beta )}$ denotes the gauge field strength
tensor. All these fields transform in the adjoint representation of the
gauge group, but we omitted the gauge index.
Latin indices are the fundamental indices of $SU(4)_R$, while the Greek
indices refer to the Lorentz group. Parameters $\eta^\alpha_I$ and
$\overline \eta^{I \dot \alpha}$ all anticommute and denote the action
of supercharges $Q^I_\alpha$ and $\overline Q_{I \dot \alpha}$.

A primary operator cannot, by definition, be expressed as a supersymmetry
transformation of some other operator. Hence, primary operators are products
of the scalar fields $\phi_{[IJ]}$, with color indices contracted without
the use of the $f$ symbol. Any antisymmetric contraction of gauge indices
indicates that the operator contains some descendent because the
supersymmetry transformations (\ref{susytransf}) contain a commutator of
scalar fields.

Short or ``chiral'' multiplets are generated by at most eight supersymmetry
transformations of a primary field, while the remaining transformations
annihilate the field. In the notation of Eq.~(\ref{susytransf}),
a short multiplet is generated by at most four $Q^I_\alpha$'s (corresponding
to $\eta^\alpha_I$'s) and no more than four $\overline Q_{I \dot \alpha}$
($\overline \eta^{I \dot \alpha}$).

Primary gauge-invariant fields $\phi^k$ in the $(0,k,0)$ representation of
$SU(4)_R$ generate short multiplets. There are several arguments in the
literature explaining that fact~\cite{oscillator,Gunaydin,Ferrara}.
These arguments do not make any reference to the gauge structure of
the primary fields. Short multiplets were constructed using the
oscillator method, in which the representations of non-compact
groups are built from unitary representations of a compact
subgroup~\cite{oscillator,Gunaydin}. Moreover, representations that
contain ``null vectors'' under the action of supersymmetry transformations
are short~\cite{review,Ferrara}. This argument also relies on the global
symmetries of the primary field, and not its gauge structure.

Another way of showing that $\phi^k$ in $(0,k,0)$ are in short multiplets
is by considering the ${\cal N}=1$ superconformal algebra. In the
${\cal N}=1$ language, only $SU(3)\times U(1)$ subgroup of $SU(4)$ is
manifest. Scalar fields are in the antisymmetric ($\bf{6}$) representation,
they decompose into chiral ${\bf 3}_{2/3}$ and antichiral
${\bf \bar{3}}_{-2/3}$ of $SU(3)\times U(1)$. The decomposition of $\phi^k$
in $(0,k,0)$ into $SU(3)\times U(1)$ involves a $k$-index symmetric tensor
of $SU(3)$ carrying the R-charge of $\frac{2 k}{3}$. Such a field must be
${\cal N}=1$ chiral since it is a product of $k$ scalar components of
chiral superfields. This means that the original multiplet must be short
and that its dimension is fixed by the ${\cal N}=1$ superalgebra to be
$k$.

\section{Two- and Three-Point Correlators}
\label{sec:corr}
\setcounter{equation}{0}
\setcounter{footnote}{0}

Conformal symmetry uniquely specifies the form of two- and three-point
correlator functions. Anomalous dimensions of operators show up in
correlators as logarithmic corrections. Such logarithms arise as the result
of expanding exponentials of the anomalous dimensions in the powers
of coupling.  In the case of protected operators, only the overall
coefficient of a correlator function can depend on the coupling constant.
We will show the absence of radiative corrections at $O(g^2)$ in any
form---containing either logarithms or powers of the space-time points.
This affirms that the operators we consider have protected dimensions.

We now show that two-point correlators of arbitrary linear combinations
of operators in the $(0,k,0)$ representation are not corrected at order
$g^2$. We denote the single-trace operators by
\begin{displaymath}
  O_k=\tr(T^{a_1} \ldots T^{a_k}) \, \phi_{I_1 J_1}^{a_1} 
      \ldots \phi_{I_k J_k}^{a_k},
\end{displaymath}
while generic multi- or single-trace operators by
\begin{displaymath}
  M_k^r=\tr(T^{a_1} \ldots T^{a_{m_1}}) \ldots \tr(T^{a_{m_r+1}} \ldots
    T^{a_k}) \, \phi_{I_1 J_1}^{a_1} \ldots \phi_{I_k J_k}^{a_k}.
\end{displaymath}
The set of numbers $m_1, \ldots, m_r$ defines the number of generators in
each of the $r$ traces. The $k$-th trace contains $m_k-m_{k-1}$ generators,
where $m_0=0$. With this definition $O_k=M_k^1$. In the above formulas,
matrices $T^a$ are the generators of any simple nonabelian gauge
group. Indices $a$ are the adjoint indices. The generators can be
in any representation since we will only use the cyclic property
of the trace and the Lie algebra relations about commutators. 

Following Ref.~\cite{ours} we work in the ${\cal N}=1$ component language.
We refer to the scalar components of chiral superfields as $z^n$ and
the antichiral ones as $\bar{z}^n$, where $n=1,2,3$ is the $SU(3)$ index.
The calculation we carry out here follows closely that
of Ref.~\cite{ours}. We will use the same notation.
We denote the propagator of scalar fields by
\begin{displaymath}
  G(x,y)=\frac{1}{4 \pi^2} \frac{1}{(x-y)^2}.
\end{displaymath}
There are two types of interactions that enter the calculation of the
two-point function. The first interaction is the self-energy correction
to the scalar propagator
\begin{equation}
\label{eq:A}
    \mbox{\raisebox{-.2truecm}{\epsfysize=0.5cm\epsffile{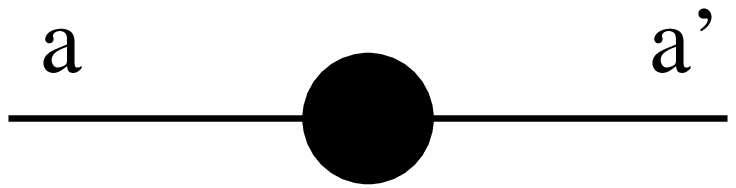} }}
    = \delta^{aa'} C_2(G) A(x,y) G(x,y),
\end{equation}
where $a, a'$ are the gauge indices, and $C_2(G)$ is the Casimir of the
adjoint representation of the gauge group $G$.
We will not need the explicit form of function
$A(x,y)$. Second, there are four-point interactions which come from
summing the gluon exchange diagram and quartic interaction originating
from the D-term. The sum of these two diagrams gives
\begin{equation}
\label{eq:B}
    \mbox{\raisebox{-.45truecm}{\epsfysize=1cm\epsffile{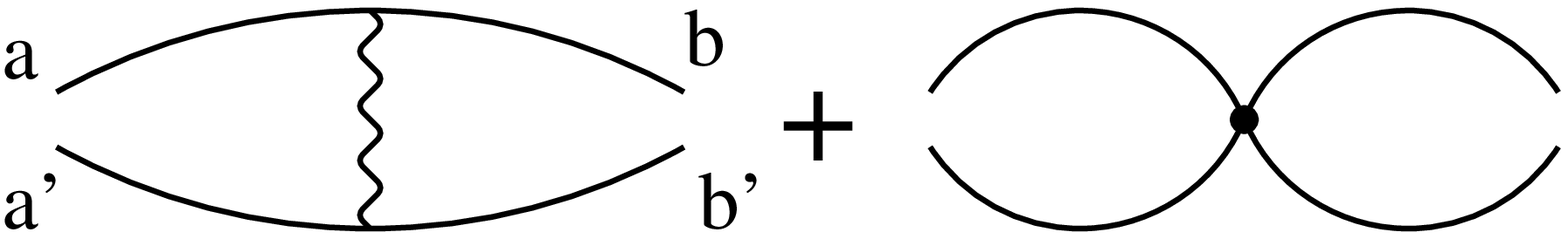} }}
  =(f^{pab} f^{p a' b'} + f^{pab'} f^{p a' b})  B(x,y) G^2(x,y),
\end{equation}
where, again, the explicit form of $B(x,y)$ is irrelevant.

We always choose the highest-weight state in the $k$-index symmetric
representation of $SU(3)$, in order to avoid complications with the
combinatorics of flavor symmetry. For a two-point function this implies
that one of the operators is proportional to $k$ powers of $z^1$, the
other to $\bar{z}^1$. The tree-level expression for the correlator of two,
not necessary identical, operators
$M_k$ is given by
\begin{eqnarray}
\label{tree}
  \langle M_k^{(1)}(x) M_k^{(2)}(y) \rangle & =& G(x,y)^k P^{(1,2)}_{k,k,0}, \\
  P^{(1,2)}_{k,k,0} &=& \sum_{\rm perms\ \sigma}
    \left[ \tr(T^{a_1} \ldots T^{a_{m_1}})
    \ldots \tr(T^{a_{m_r+1}} \ldots T^{a_k})\right] \nonumber \\
  && \cdot \left[ \tr(T^{a_{\sigma(1)}} \ldots T^{a_{\sigma(m'_1)}}) \ldots  
    \tr(T^{a_{\sigma(m'_{r'}+1)}} \ldots T^{a_{\sigma(k)}})\right]. \nonumber
\end{eqnarray}
The self-energy graphs are trivial to incorporate: we need to multiply one of
the propagators by $C_2(G) A(x,y)$ and then sum over all propagators.
This sum is equal to
\begin{equation}
\label{2p-A}
  k\, C_2(G) A(x,y) G(x,y)^k P^{(1,2)}_{k,k,0}.
\end{equation}
We now insert the interaction of Eq.~(\ref{eq:B}) into the tree-level
expression (\ref{tree}) and sum over all pairs of lines. The $f$ symbols
are converted into commutators on one of the multi-trace expressions.
We do not repeat the whole expression (\ref{tree}) with additional
commutators inserted, but instead write only the relevant part
\begin{equation}
  \sum_{i\neq j=1}^k
  \left\{ \tr(T^{a_1} \ldots [T^{a_i},T^p] \ldots T^{a_{m_1}}) \ldots
         \tr(\ldots [T^{a_j},T^p] \ldots )
    \ldots \tr(T^{a_{m_r+1}} \ldots T^{a_k})\right\}.
\end{equation}
The sum over all pairs of indices can be expressed as
\begin{equation}
\label{sumsplit}
  \sum_{i \neq j=1}^k = \sum_{\stackrel{\scriptstyle{i,j \in \rm different}}
                                       {\scriptstyle{\rm \;\;\; traces}}} +
      \sum_{\stackrel{\scriptstyle{\rm each}}{\scriptstyle{ \rm  trace}}} \;\;
      \sum_{\stackrel{\scriptstyle{i\neq j \in {\rm same}}}
                     {\scriptstyle{\rm \hspace{7mm} trace}}}
\end{equation}
We now use the identity
\begin{equation}
\label{zero}
  \sum_{i=1}^k (M_1 \ldots [M_i,N] \ldots M_k)=0
\end{equation}
to observe that 
\begin{equation}
\label{difftraces}
  \sum_{\stackrel{\scriptstyle{i,j \in \rm different}}
  {\scriptstyle{\rm \;\;\; traces}}} \left\{ \tr(T^{a_1} \ldots 
  [T^{a_i},T^p] \ldots T^{a_{m_1}}) \ldots \tr(\ldots [T^{a_j},T^p] \ldots )
  \ldots \right\}  =0.
\end{equation}
This is the observation that allows us to extend the results of
Ref.~\cite{ours} from single traces to multiple traces. We will use similar
identities repeatedly. The remaining sum in Eq.~(\ref{sumsplit}) can be
performed for each trace individually. We can again use Eq.~(\ref{zero})
to rewrite the sums within a given trace
\begin{eqnarray}
\label{eachtrace}
  && \sum_{i \neq j=1}^{k_i} \tr(\ldots [T^i,T^p] \ldots [T^j,T^p] \ldots )=
        \nonumber \\
  &&-\sum_{i=1}^{k_i} \tr (\ldots [[T^i,T^p],T^p] \ldots) =
    - k_i C_2(G) \tr (\ldots),
\end{eqnarray}
where $k_i$ is the number of generators in the trace.
It is now easy to perform the sum over each trace, which gives
\begin{equation}
\label{2p-B}
  \frac{1}{2} k B(x,y) G(x,y)^k P^{(1,2)}_{k,k,0},
\end{equation} 
where the extra factor of $1/2$ is a combinatoric factor that prevents
overcounting of diagrams. The additional minus sign comes from converting
two $f$ symbols into commutators. It is now apparent that the cancellation
of radiative corrections happens for the same reason it does in the
two-point function of single traces. It was shown in Ref.~\cite{ours}
that $2 A+B=0$ due to a non-renormalization theorem
for $\langle O_2 O_2 \rangle$, which is a consequence of $O_2$ being in
the multiplet containing the energy-momentum tensor.

Similarly, we can repeat the analysis for the
$\langle M_k(x) M'_k(y) O_2(w) \rangle$ correlator. Calculation for the
general three-point functions will be presented in
the Appendix. In addition to $O(g^2)$ interactions of Eqs.~(\ref{eq:A}) and
(\ref{eq:B}) there is a four-scalar interaction involving three space-time
points. This can be illustrated as
\begin{equation}
\label{eq:C}
    \mbox{\raisebox{-.45truecm}{\epsfysize=1.2cm\epsffile{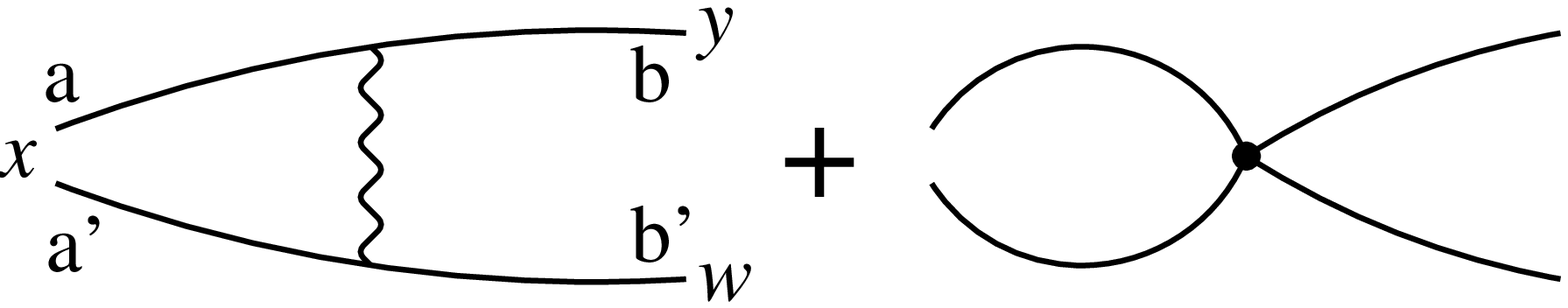} }}
 =(f^{pab} f^{p a' b'} + f^{pab'} f^{p a' b}) C(x;y,w) G(x,y) G(x,w),
\end{equation}
where the form of function $C$ will not be important for the calculation.
In some cases only the $D$-term quartic interaction contributes in the above
diagram defining $C$. Whenever both $F$- and $D$-terms contribute,
we denote the corresponding function by $C'$.

The tree-level expression for this correlator is
\begin{eqnarray}
  \langle M_k^{(1)}(x) M_k^{(2)}(y) O_2(w) \rangle&= &
          G(x,y)^{k-1} G(x,w) G(y,w)
                             P^{(1,2)}_{k,k,2} t_{11}, \nonumber \\
  P^{(1,2)}_{k,k,2} &=& k P^{(1,2)}_{k,k,2},
\end{eqnarray}
where $t_{ij}$ is an $SU(3)$ flavor matrix. This matrix enters our definition
of $O_2=\bar{z}^i  t_{ij} z^j$. In the above equation $M_k^{(1)}=(z^1)^k$
and  $M_k^{(2)}=(\bar{z}^1)^k$. The additional factor of $k$ comes from
distinguishing one line in each multi-trace expression. It is straightforward
to sum all possible self-energy diagrams since they multiply each propagator
by the same expression. The sum is proportional to the tree-level term,
which we omit, times the following factor
\begin{equation}
  C_2(G) \left[(k-1) A(x,y) + A(x,w) + A(y,w)\right].
\end{equation}

Next, we include quartic interactions (\ref{eq:B}) by summing over all pairs
of $k-1$ propagators that connect the operators $M_k^{(1)}(x)$ and
$M_k^{(2)}(y)$. Repeating the same steps we have already explained for
the two-point function, we convert the $f$ symbols into commutators and
then perform the sum over pairs of indices. Since there is only one
distinguished generator in the traces at points $x$ and $y$,
Eq.~(\ref{difftraces}) still holds. The distinguished generator is the
one connected by a propagator to the operator $O_2$.
One can always use the identity (\ref{zero}) for a trace that
does not contain the distinguished line. The remaining sum in
Eq.~(\ref{sumsplit}) is easy to carry out. For the traces without
the distinguished generator we pick a factor of $-k_i$, which is the
negative of the number of elements in a given trace, just like
we did in Eq.~(\ref{eachtrace}). For the trace with the distinguished
generator, the algebra is slightly more involved. We can always use
the cyclic property of the trace to place the special generator as
the last one in the trace. Then, we use the following identity
\begin{eqnarray}
\label{rainbow2kk}
  && \sum_{i\neq j=1}^{k_i-1} \tr(\ldots [T^{a_i},T^p]\ldots [T^{a_j},T^p]
     \ldots) \nonumber \\
  &=&  -\sum_{i=1}^{k_i-1}\left\{ \tr(\ldots [[T^{a_i},T^p],T^p] \ldots )+ 
           \tr(\ldots [T^{a_i},T^p] \ldots [T^{a_{k_i}},T^p]) \right\} 
     \nonumber \\
  &=&  - (k_i-1-1)C_2(G) \tr(\ldots)
\end{eqnarray}
to complete the evaluation of the sum over pairs of propagators between
the two operators $M_k$'s. There are $k-1$ such propagators.
Due to the additional $-1$ in Eq.~(\ref{rainbow2kk})
the final expression is proportional to $k-2$.

There are three possible ways of inserting interaction (\ref{eq:C}) into the
tree-level expression, depending on which vertex the two lines grouped
together enter. This distinguished vertex in Eq.~(\ref{eq:C}) corresponds
to the space-time point $x$. When the distinguished vertex is one of the
vertices corresponding to operators $M_k^{(1)}(x)$ or $M_k^{(2)}(y)$,
we need to sum over all propagators connecting these vertices.
This single sum is easy to perform using the identity (\ref{zero})
since the sum over all, but one, traces vanishes.
The selected trace is the one which involves a generator linked to the
operator $O_2(w)$ by the scalar propagator. The algebra is identical to
the one we used in the previous equation, which is
\begin{equation}
\label{legrainbow2kk}
  \sum_{i=1}^{k_i-1} \tr(\ldots [T^{a_i},T^p] \ldots [T^{a_{k_i}},T^p]) =
            -C_2(G) \tr(T^{a_1}\ldots T^{a_{k_i}}).
\end{equation}
When Eq.~(\ref{eq:C}) is inserted with the distinguished vertex corresponding
to the operator $O_2(w)$ there is no need to perform any sums since there are
only two propagators entering this vertex. However, the choice of flavors at
this vertex is such that both the $D$-term and $F$-term vertices contribute
in this case. Thus, the expression is proportional to $C'(w;x,y)$ instead of
$C$. The sum of all $O(g^2)$ contributions is equal to the Born term times
\begin{eqnarray}
\label{finalsum2kk}
  & C_2(G)&\big[ (k-1) A(x,y) + A(x,w) + A(y,w) + \frac{k-2}{2} B(x,y)+ 
     \nonumber \\
  && 2 C(x;y,w) + 2 C(y;x,w)+2 C'(w;x,y)\big].
\end{eqnarray}
The minus sign in front of functions $C$ compared to Eq.~(\ref{legrainbow2kk})
is the result of converting $f$ symbols into commutators. The factor of two
comes from two pairs of $f$ symbols in the definition (\ref{eq:C}) of
function $C$. Eq.~(\ref{finalsum2kk}) is equal to zero as a result of
non-renormalization of the $\langle O_2(x) O_2(y) O_2(w)\rangle$
correlator, which implies that \cite{ours}
\begin{equation}
\label{AplusC}
  A(x,y) + A(x,w) + A(y,w) + 2 [ C(x;y,w) + C(y;x,w)+C'(w;x,y)]=0.
\end{equation}
This equation together with the fact that $2 A+B=0$ imply the lack of $O(g^2)$
corrections to the $\langle M_k(x) M'_k(y) O_2(w) \rangle$ correlator.
In the Appendix, we show that $O(g^2)$ corrections for general three-point
functions $\langle M_{k_1} M_{k_2} M_{k_3}\rangle$ vanish.
\section{Conclusions}
\setcounter{equation}{0}
\setcounter{footnote}{0}

We investigated correlators of primary fields of short multi-trace operators
in ${\cal N}=4$ supersymmetry. Such primary fields contain $k$ scalar fields
in the $(0,k,0)$ representation of $SU(4)_R$ and arbitrary symmetric
contraction of color indices. We have shown that two- and three-point
functions of chiral primary fields do not receive radiative corrections at
order $g^2$ in perturbation theory. This result is valid for any gauge group
due to the non-renormalization theorem for two- and three-point correlators
of the so-called ``massless'' multiplet, which includes the energy-momentum
tensor. The result is also independent of the the size of the gauge group,
so it is not a leading $N$ result in the large-$N$ expansion.

Our results suggest that two- and three-point correlators of arbitrary short  
multiplets are not renormalized. If these results hold beyond low orders in
perturbation theory, it indicates that there must be a symmetry reason for
such non-renormalization. The so-called bonus $U(1)_Y$ symmetry~\cite{bonus1}
explains why correlators of operators which correspond to single-particle
states of supergravity on AdS spaces are not renormalized.
Perhaps it is possible to extend the arguments about the $U(1)_Y$ symmetry
to correlators of short operators corresponding to multi-particle states.
Since all the quantum numbers of the single- and multi-trace chiral
primary operators are identical, non-renormalization theorems
are likely to be a consequence of the full ${\cal N}=4$ superconformal
symmetry. Superconformal Ward identities would not distinguish between
single-trace and multi-trace operators.

It would be interesting to explore the implications of these results
for the duality between supergravity and conformal field theory.
One of the issues that require better understanding is the correspondence
between operators in the conformal field theory and supergravity states.
It is believed that single-particle states of supergravity correspond to
short single-trace operators. However, the single- and multi-trace
operators mix. It is easy to check that at the tree-level, the two-point
functions $\langle O_k M_k\rangle$ do not vanish. In the $SU(N)$ theory,
two-point functions of single-trace operators $O_k$ are proportional to
$N^k$ in the leading order in $N$. Likewise, in the leading order
$\langle O_k M_k^r \rangle\propto N^{k-r+1}$, where $r$ is the number
of color traces in the operator $M_k^r$. Thus, the mixing of operators
$O_k$ with $M_k$'s is suppressed in the large-N limit.

\section*{Acknowledgements}
I would like to thank Dan Freedman, Ken Intriligator and Aneesh Manohar
for discussions and comments on the manuscript. I am especially indebted to
Dan Freedman for the correspondence that started this project and for
encouraging me to write up the results. 
This work was supported by the U.S. Department of Energy under
Contract DOE-FG03-97ER40506.

\appendix
\renewcommand{\thesection}{Appendix}
\section{$\langle M_{k_1} M_{k_2} M_{k_3} \rangle$}
\renewcommand{\theequation}{\Alph{section}.\arabic{equation}}
\setcounter{equation}{0}
\setcounter{footnote}{0}

We now extend the calculations of Section~\ref{sec:corr}. We show that
arbitrary three-point correlators
$\langle M_{k_1}(x) M_{k_2}(y) M_{k_3}(w) \rangle$ are not corrected at
order $g^2$. We define the sum over all permutations of indices
in a given multi-trace expression as
\begin{equation}
  M_{k_i} \propto \str(T^{a_1} \ldots T^{a_{k_i}}) 
  \equiv \frac{1}{k_i !}
  \sum_{\rm{perms}\ \sigma}
  \tr(T^{a_{\sigma(1)}} \ldots T^{a_{\sigma(m_1)}}) \ldots 
  \tr(T^{a_{\sigma(m_r+1)}} \ldots T^{a_{\sigma(k_i)}}),
\end{equation}
where the number of traces and the number of generators in each trace
depends on the gauge structure of a particular operator. The number of
propagators in the tree-level expression that connect operators with $k_1$
and $k_2$ fields are $\alpha_3=(k_1+k_2-k_3)/2$ and so on. As in the previous
sections, $M_{k_1} \propto (z^1)^{k_1}$ is ${\cal N}=1$ chiral and it is
the highest-weight field of the completely symmetric $SU(3)$ representation.
Similarly, $M_{k_2} \propto (\bar{z}^1)^{k_2}$ is antichiral, while
$M_{k_3} \propto (z^1)^{\alpha_1} (\bar{z}^1)^{\alpha_2}$.

With the above definitions, the tree-level expression for the correlator is 
\begin{eqnarray}
   \langle M_{k_1}(x) M_{k_2}(y) M_{k_3}(w) \rangle & \propto &
    G(x,y)^{\alpha_3} G(x,w)^{\alpha_2} G(y,w)^{\alpha_1} P_{k_1,k_2,k_3},
      \nonumber \\
   P_{k_1,k_2,k_3}&=&\str (T^{a_1}\ldots T^{a_{\alpha_3}} T^{b_1} \ldots
                           T^{b_{\alpha_2}}) \,
   \str (T^{b_1}\ldots T^{b_{\alpha_2}} T^{c_1} \ldots T^{c_{\alpha_1}})
   \nonumber \\
  &&  \cdot \str (T^{c_1}\ldots T^{c_{\alpha_1}} T^{a_1} \ldots
                  T^{a_{\alpha_3}}).
\end{eqnarray}
In the above formula we omitted the $SU(3)$ flavor generator,
which contracts the flavor indices of chiral and antichiral fields in
$M_{k_3}(w)$. The computation of radiative corrections is very similar
to the one we presented for the $\langle M_k M'_k O_2 \rangle$ correlator,
only the algebra of traces is more complicated. There are three types
of diagrams that contribute to the $O(g^2)$ corrections. These correspond to
insertions of $O(g^2)$ interactions of Eqs.~(\ref{eq:A}), (\ref{eq:B}),
and (\ref{eq:C}) into the tree-level expression for the correlator.

The self-energy graphs are straightforward to incorporate. They give a factor
of $C_2(G) A$ for every propagator, so we get
\begin{equation}
\label{3p-A}
  C_2(G)\left[ \alpha_3 A(x,y) + \alpha_2 A(x,w) + \alpha_1 A(y,w) \right]
\end{equation}
times the Born expression. Next, we include the interactions
of Eq.~(\ref{eq:C}). For each of the three vertices we need to evaluate
\begin{equation}
\label{Wk1k2k3}
  W_{k_1,k_2,k_3} \equiv
   \sum_{i=1}^{\alpha_3} \sum_{j=1}^{\alpha_2}
      \str (T^{a_1} \ldots [T^{a_i},T^p]
      \ldots T^{a_{\alpha_3}} T^{b_1} \ldots [T^{b_i},T^p] 
      \ldots T^{b_{\alpha_3}}) 
      \, \str (\ldots) \, \str (\ldots). 
\end{equation}
However, it turns out that all three such expressions are identical. We can
do the algebra for a fixed, yet sufficiently general, permutation of the
generators. Performing the sums over all permutations will not affect the
result. We divide the traces in the multi-trace expression
$\str(T^{a_1}\ldots T^{b_{\alpha_3}})$ into three categories: traces
containing only generators with index $a$, only index $b$, and containing
both indices. For simplicity, we choose the permutations of indices such
that for traces with two different indices, the generators with index $a$
come before the ones with index $b$. For example,
\begin{eqnarray}
\label{fixedperm}
 && \tr(T^a \ldots T^a) \ldots \tr(T^a \ldots T^a)
    \tr(T^a \ldots T^a T^b \ldots T^b)
    \ldots \tr(T^a \ldots T^a T^b \ldots T^b) \nonumber \\
 && \cdot \tr(T^b \ldots T^b) \ldots \tr(T^b \ldots T^b).
\end{eqnarray}
We choose analogous permutations for the other two traces. When performing
the sums $\sum_{i=1}^{\alpha_3}$ and $\sum_{j=1}^{\alpha_2}$ in
Eq.~(\ref{fixedperm}), we can neglect all traces with one kind of
indices only. Sums over these traces vanish due to Eq.~(\ref{zero}).
Our goal is to show that we can move the sums from one multi-trace
expression to another. Converting the commutators into the $f$ symbols
and then back into commutators on another trace, we get
\begin{eqnarray}
\label{movesums}
  W_{k_1,k_2,k_3}&=&\str (T^{a_1}\ldots T^{b_{\alpha_3}}) 
     \sum_{j=1}^{\alpha_2} 
     \str (T^{b_1}\ldots [T^{b_i},T^p]
     \ldots T^{b_{\alpha_2}} T^{c_1} \ldots T^{c_{\alpha_1}})
     \nonumber \\
  && \cdot \sum_{i=1}^{\alpha_3} \str (T^{c_1}\ldots T^{c_{\alpha_1}}
     T^{a_1} \ldots [T^{a_i},T^p] \ldots T^{a_{\alpha_3}}).
\end{eqnarray}
Within a given multi-trace expression we can convert a sum over one kind
of index into minus the sum over another index
\begin{eqnarray}
 && \sum_{i=1}^{\alpha_3} \tr(T^{a_1} \ldots  [T^{a_i},T^p]
     \ldots T^a T^b \ldots T^b) \ldots
     \tr(T^a \ldots T^{a_{\alpha_3}} T^b \ldots T^b) \\
 &=& - \sum_{j=1}^{\alpha_2} \tr(T^a \ldots T^a T^{b_1}
    \ldots [T^{b_j},T^p] \ldots T^b)
    \ldots \tr (T^a \ldots T^a T^b \ldots T^{b_{\alpha_2}}) \nonumber
\end{eqnarray} 
because we can use Eq.~(\ref{zero}) within each trace independently.
In Eq.~(\ref{movesums}) we can exchange the sum over $b$'s into the sum
over $c$'s using the above trick. We can then move the sum over $c$'s into
the last trace in Eq.~(\ref{movesums}) by converting $f$ symbols into traces.
This shows that the expression defined in Eq.~(\ref{Wk1k2k3}) is
universal for all insertions of interactions (\ref{eq:C}), hence we obtain
\begin{equation}
\label{3p-C}
  -2 W_{k_1,k_2,k_3} \left[ C(x;y,w)+C(y;x,w)+C'(w;x,y)\right].
\end{equation}
The overall factor of $-2$ is the result of converting two pairs of $f$
symbols into propagators. This is the same factor as the one appearing
in Eq.~(\ref{finalsum2kk}).

The final step is evaluating the contributions of interactions of
Eq.~(\ref{eq:B}). For each set of indices $a$, $b$, and $c$, we need
to compute the double sum
\begin{eqnarray}
  &&\sum_{i\neq j=1}^{\alpha_3} 
    \str (T^{a_1}\ldots [T^{a_i},T^p] \ldots [T^{a_j},T^p] \ldots
          T^{a_{\alpha_3}}
          T^{b_1} \ldots T^{b_{\alpha_2}})
    \str(\ldots ) \str(\ldots ) \nonumber \\
  &=& - \Big[ \sum_{i=1}^{\alpha_3} \sum_{j=1}^{\alpha_2}
     \str (T^{a_1}\ldots [T^{a_i},T^p] \ldots T^{a_{\alpha_3}} T^{b_1} \ldots
        [T^{b_j},T^p] \ldots  T^{b_{\alpha_2}}) \nonumber \\
  && \hspace{5mm}+ \sum_{i=1}^{\alpha_3}
                  (T^{a_1}\ldots [[T^{a_i},T^p],T^p] \ldots 
     T^{b_1} \ldots T^{b_{\alpha_2}}) \Big] \str(\ldots ) \str(\ldots )
     \nonumber \\
  &=& -C_2(G) \alpha_3 P_{k_1,k_2,k_3} - W_{k_1,k_2,k_3},
\end{eqnarray}
where the first equality follows from applying Eq.~(\ref{zero}) to a fixed
permutation of indices, just like we did in Eq.~(\ref{fixedperm}). Therefore,
the contribution of interactions from Eq.~(\ref{eq:B}) is proportional to
\begin{eqnarray}
\label{3p-B}
  & \frac{1}{2} C_2(G) P_{k_1,k_2,k_3} [\alpha_3 B(x,y)+
    \alpha_2 B(x,w)+\alpha_1 B(y,w)] &
    \nonumber \\
  & +  \frac{1}{2} W_{k_1,k_2,k_3} [B(x,y)+B(x,w)+B(y,w)], & 
\end{eqnarray}
where the factor of $\frac{-1}{2}$ has the same origin as in Eq.~(\ref{2p-B}).
The sum of all $O(g^2)$ interactions presented in Eqs.~(\ref{3p-A}),
(\ref{3p-C}), and (\ref{3p-B}) is
\begin{eqnarray}
  && W_{k_1,k_2,k_3} \left\{ -2\left[C(x;y,w)+C(y;x,w)+C'(w;x,y)\right] 
     + \frac{1}{2} \left[ B(x,y)+B(x,w)+B(y,w) \right] \right\}  \nonumber \\
  && +C_2(G) P_{k_1,k_2,k_3} \left\{ \alpha_3 (A+\frac{1}{2}B)(x,y) + 
     \alpha_2 (A+\frac{1}{2}B)(x,w)+ \alpha_1 (A+\frac{1}{2}B)(y,w) \right\}, 
\end{eqnarray}
where the first line vanishes after using both $2 A+B=0$
and Eq.~(\ref{AplusC}).


\end{document}